\documentstyle[aps,multicol,psfig]{revtex}
\input epsf
\def\deltat{\tau}  
\def\inte{g} 	   
\def\numcop{{n}}
\def\disfac{\chi}
\def\dmhrs{{{\overline{\cal D}}^{\possym}} \Omega }

\def \ofield#1{ \Omega(\hat #1)}
\def\HRS{{\rm HRS}}

\def\action{{\cal S}_{n}}

\newcommand{\volfrac}{\varphi}	

\newcommand{\oofield} [2] {\Omega({\hat #1}_{#2})}

\def\possym{\dagger}

\def\sumin{\sum_{j=1}^{N}}

\def\rmd{d}

\def\rmi{i}

\def\wpeps{{\varepsilon}}
\def\bft{{\bf t}}
\def\nbft{{t}}
\def\aparam{{\bar{a}}}
\def\bparam{{\bar{b}}}
\def\FTav{{\cal S}}
\newcommand{\eqbreak}{
\end{multicols}
\widetext
\noindent
\rule{.48\linewidth}{.1mm}\rule{.1mm}{.1cm}
}
\newcommand{\eqresume}{
\noindent
\rule{.52\linewidth}{.0mm}\rule[-.1cm]{.1mm}{.1cm}\rule{.48\linewidth}{.1mm}
\begin{multicols}{2}
\narrowtext
}
\begin{document} 
\draft 
\title{Correlations Near the Vulcanization 
Transition: A Renormalization-Group Approach}
\author{Weiqun Peng and Paul M.~Goldbart} 
\address{Department of Physics, 
University of Illinois at Urbana-Champaign, 
1110 West Green Street, 
Urbana, Illinois 61801, U.S.A.}
  \date{December 23, 1999}
\maketitle
\begin{abstract} 
Correlators describing the vulcanization transition are 
constructed and explored via a renormalization group approach. 
This approach is based on a minimal model that accounts for 
the thermal motion of constituents and the quenched 
random constraints imposed on their motion by crosslinks. 
Critical exponents associated with the correlators are obtained 
near six dimensions, and found to equal those governing 
analogous entities in percolation theory.  Some expectations 
for how the vulcanization transition is realized in two
dimensions, developed with H.~E.~Castillo, are discussed.  
\end{abstract}
%
%
%
\begin{multicols}{2}
\narrowtext
\noindent
{\it Introduction\/}.  
The vulcanization transition (VT) is the equilibrium phase transition 
from a liquid state of matter to an amorphous solid state.  It 
occurs when a sufficient density of permanent random constraints
(e.g.~chemical crosslinks)---the quenched randomness---are introduced
to connect the constituents (e.g.~macromolecules), whose locations are
the thermally fluctuating variables.  Whilst a rather detailed 
description of the VT has emerged over the past few years within the 
context of a mean-field
approximation~\cite{prl_1987,REF:EPLandAdvPhy}, the picture of
this transition beyond the mean-field (MF) level is less certain.  

The true (i.e.~beyond-MF) critical properties of the VT (and the 
related chemical-gelation transition) have mostly been studied 
through approaches based on gelation/percolation
perspectives~\cite{REF:FloryBook,REF:PGDGbook,REF:LubIsaac}, in 
which the VT is directly or indirectly identified with the 
percolation transition.  These approaches are associated with 
{\it one single ensemble\/}, accounting either for the quenched 
disorder or for the equilibrium thermal configurations (whose change 
in character mark the transition).  Given that an essential aspect 
of the VT is the impact of the quenched random constraints on the 
thermal motion of the constituents, approaches based on a single 
ensemble cannot directly account for the effects of both types of 
fluctuations and are, thus, not entirely satisfactory.

The purpose of the present Letter is to apply renormalization-group (RG) 
ideas to a model in which both the quenched randomness and the thermal
fluctuations of the constituents are naturally and directly incorporated. 
Our analysis provides a more complete way of obtaining the critical 
exponents, and therefore sheds light on the connection between 
gelation/percolation physics and the VT.  A detailed account of this 
work can be found in Ref.~\cite{REF:PandG:VTL_pre}.

Our approach to the VT is based on a minimal Landau-Wilson effective 
Hamiltonian, which was constructed in Ref.~\cite{univ} and also shown 
to recover (at the mean-field level) the description of both the liquid 
and emergent amorphous solid states, known earlier from the analysis of 
various semi-microscopic models~\cite{REF:EPLandAdvPhy,REF:endlink}.  
The order parameter for 
the VT is a function that encodes both the gel fraction $q$ (i.e.~the 
fraction of monomers localized) and the distribution of localization 
length of the localized monomers (as well as other diagnostics).  
Support for the mean-field picture of the solid state has emerged 
from extensive molecular dynamics computer simulations of 
three-dimensional, off-lattice, interacting, macromolecular systems, 
due to Barsky and Plischke~\cite{REF:SJB_MP}. 
\par
\noindent
{\it Modeling the VT---the order parameter\/}.  
The appropriate (dimensionless) order parameter for the VT, capable 
{\it inter alia\/} of distinguishing between the liquid and amorphous 
solid states of randomly crosslinked macromolecular systems (RCMSs), 
is the following function of $\numcop$ wave-vectors 
$\{{\bf k}^1,\ldots,{\bf k}^{\numcop}\}$:
\begin{equation}
\Big[\,
\frac{1}{N} \sumin \int_{0}^{1}ds\,
\big\langle e^{i{\bf k}^{1}\cdot{\bf c}_{j}(s)} \big\rangle_{\disfac}
\cdots
\big\langle e^{i{\bf k}^{\numcop}\cdot{\bf c}_{j}(s)} \big\rangle_{\disfac}
\,\Big], 
\label{EQ:opDefinition}
\end{equation}
where $N$ is the total number of macromolecules, ${\bf c}_{j}(s)$
(with $j=1,\ldots, N$ and $0\leq s\leq 1$) is the position in
$d$-dimensional space of the monomer at fractional arclength $s$ along
the $j^{\rm th}$ macromolecule, $\langle\cdots\rangle_{\disfac}$ denotes a
thermal average for a particular realization $\disfac$ of the quenched
disorder (i.e.~the crosslinking), and $\left[\cdots\right]$ represents
a suitable averaging over this quenched disorder.  It is worth
emphasizing that the disorder resides in the specification of what
monomers are crosslinked together: the resulting constraints do
not explicitly break the translational symmetry of the system.
\par
\noindent
{\it Modeling the VT---the minimal model\/}.  
Following Deam and Edwards~\cite{REF:DeamEd}, we adopt a constraint 
distribution appropriate to the situation of instantaneous 
crosslinking of the equilibrium melt or solution, and invoke the
replica trick to incorporate the consequences of the permanent random
constraints.  Thus, we are led to the need to work with the 
$\numcop \to 0$ limit of systems of $\numcop + 1$ replicas. The 
additional (i.e.~zeroth) replica incorporates the constraint 
distribution.

The form of the minimal model can be determined, in the spirit of 
the Landau approach, from the nature of the order parameter and the 
symmetries of the effective (i.e.~pure but replicated) Hamiltonian, 
along with the assumptions of the analyticity of this Hamiltonian 
and the continuity of the transition~\cite{univ}. This scheme leads 
to the following minimal model, which takes the form of a cubic field 
theory involving the order parameter field $\ofield{k}$ living on 
$(n+1)$-fold replicated $d$-dimensional space:
\vspace*{-0.01in}
\begin{mathletters}
\begin{eqnarray}
&&f 
\propto
\lim_{n\to 0}n^{-1}\ln\big[Z^{n}\big]\,,
\quad
\left[Z^{n}\right] 
\propto
\int \dmhrs 
\exp( - \action)\,, 
\label{EQ:Partition}
\\
&&\action
\big(\{\Omega\}\big)
=
N\sum_{\hat{k} \in {\HRS}}
\Big(-\aparam\deltat+\frac{\bparam}{2}|\hat{k}|^2\Big)
\big\vert\ofield{k}\big\vert^{2}
\nonumber
\\
&&\quad -N\inte
\sum_{\hat{k}_1,\hat{k}_2,\hat{k}_3\in\HRS}
\oofield{k}{1}\,
\oofield{k}{2}\,
\oofield{k}{3}\,
\delta_{{\hat{k}_1}+{\hat{k}_2}+{\hat{k}_3},{\hat{0}}}\,.
\label{EQ:LG_longwave}
\end{eqnarray}
\end{mathletters}
Here, $\deltat$ is the control parameter, which measures 
the reduced density of random constraints, and the coefficients $\aparam$, 
$\bparam$ and $\inte$ encode the microscopic details of the system.  
We denote averages weighted with 
$\exp(-\action)$ by $\langle\cdots\rangle^{\FTav}$, use the symbol 
${\hat k}$ to denote the replicated wave-vector 
$\{{\bf k}^0, {\bf k}^1,\ldots, {\bf k}^n\}$, 
and define the extended scalar product
$\hat{k}\cdot\hat{c}$ by 
${\bf k}^0\cdot{\bf c}^0
+{\bf k}^1\cdot{\bf c}^1
+\cdots
+{\bf k}^n\cdot{\bf c}^n$. The symbol $\hat{k} \in {\HRS}$
indicates that a summation over replicated wave-vectors is restricted
to those containing at least two nonzero component-vectors 
${\bf k}^\alpha$. (We say that this kind of wave-vector lies in
higher-replica-sector, i.e.~the HRS.)\thinspace\ This condition reflects 
the important fact that no crystalline order (or any other kind of 
macroscopic inhomogeneity) is present in the vicinity of the VT; it 
changes the symmetry of the effective Hamiltonian and, hence, can 
(and in fact does) play a crucial role.  Consequently, the functional 
integral is to be performed over fields obeying the linear constraint
of lying in the HRS.
\par
\noindent 
{\it Ginzburg criterion\/}. 
To estimate the range of $\deltat$ about the critical value, within 
which the effects of order-parameter fluctuations are relatively 
strong, we follow the conventional strategy of constructing a loop 
expansion (in the present setting, an expansion in the inverse monomer 
density) for the 2-point vertex function to one-loop order, examining 
its low-wave-vector limit, thus obtaining the inverse susceptibility 
$\Xi$.
 \begin{figure}[hbt]
 \vskip0.20cm
 \epsfxsize=2in
 \centerline{\epsfbox{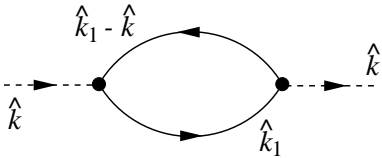}} 
 \vskip0.20cm
 \caption{One-loop correction to the 2-point vertex function.  
 \label{FIG:ginzburg}}
 \end{figure}
The only one-loop correction comes from the diagram shown in
Fig.~\ref{FIG:ginzburg}; thus, we find
\begin{equation}
\frac{1}{N\Xi}
\approx
	-2\aparam(\deltat-\deltat_{\rm c}) 
	\left( 1- 18 \frac{V\inte^2 J_d}{N\bparam^{d/2}} 
	(-2\aparam\deltat)^{(d-6)/2} \right), 
\label{EQ:Ginz_Crit}
\end{equation}

where $\deltat_{\rm c}>0$ is the critical value of $\deltat$, 
shifted due to the inclusion
of one-loop corrections, and $J_d$ is an unimportant dimensionless
number.  Equation~(\ref{EQ:Ginz_Crit}) shows that the upper critical
dimension for the VT is six. To determine the physical content of the
Ginzburg criterion, we invoke the values of the coefficients
appropriate for the semi-microscopic model of RCMSs, and find the
following form of the Ginzburg criterion: for $d<6$, fluctuations
cannot be neglected for values of $\deltat$ satisfying:
\begin{equation}
\vert\deltat\vert
\lesssim 
\left({L/{\ell}}\right)^{-\frac{d-2}{6-d}}
\left({\volfrac/{\inte^{2}}}\right)^{-{2}/(6-d)}, 
\label{EQ:GZinRCMS}
\end{equation}
Here, $L$ is the arclength of each macromolecule, $\ell$ the
persistence length, and $\volfrac\equiv(N/V)(L/\ell)\ell^{d}$ the
volume fraction.  Equation~(\ref{EQ:GZinRCMS}) shows that the
fluctuation-dominated regime is narrower for longer macromolecules and
higher densities (for $2<d<6$).  Such dependence on the degree of
polymerization $L/\ell$ is precisely that argued for by de~Gennes on
the basis of a percolation picture~\cite{REF:DeGennes}.
\par
\noindent 
{\it Epsilon expansion\/}. 
We now apply the $\wpeps$ ($\equiv 6-d$) expansion to derive the RG
flow equations near the upper critical dimension, and discuss the
resulting fixed-point structure and critical exponents.  In order to
streamline the presentation, we may, by suitably redefining the scales of
$\ofield{k}$ and $\hat{k}$ in Eq.~(\ref{EQ:LG_longwave}), absorb the
coefficients $\aparam$ and $\bparam$ (i.e.~set $\aparam=1$, $\bparam=1$).
 \begin{figure}[hbt]
 \epsfxsize=2.0truein 
 \centerline{\epsfbox{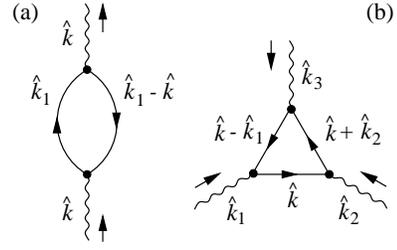}} 
 \vskip0.50cm
 \caption{Contributing one-loop diagrams.  Full lines indicate bare 
 \HRS\ correlators for short-wavelength fields (i.e.~fields lying in 
 the momentum shell); wavy lines indicate long-wavelength fields.
 \label{FIG:renorm}}
 \end{figure}%
We shall be working to one-loop order and, correspondingly, the
contributing diagrams are those depicted in Figs.~\ref{FIG:renorm}a,b. 
After taking proper care of the constraints on the wave-vector 
summations, and passing to the $n \to 0$ limit, the flow 
equations are (with higher order terms omitted) found to be
\begin{mathletters}
\begin{eqnarray}
d\deltat/d\ln b
&=& 
	2\,\deltat -C_0\, \inte^2- C_0^{\prime}\,\deltat\, \inte^2 - 
	C_1\, \deltat \,\inte^2,
\label{EQ:ctmflowdel}
\\
d\inte/d\ln b
&=& 
	\inte(\wpeps/2 - C_3\inte^2 -\frac{3}{2}C_1 \inte^2),
\label{EQ:ctmflowint}
\\
dz/d\ln b
&=& 
	{1\over{2}}(d+2-C_{1}\inte^{2}),
\label{EQ:ctmflowzee}
\end{eqnarray}
\end{mathletters}
where $b$ is the length-rescaling factor, 
$z$ is the field-rescaling factor, and 
the (constant) coefficients in the flow equations are given by 
$(C_0, C_0^{\prime}, C_1, C_3)
= ({V}/{N}) ({S_6}/{(2\pi)^6})
( 9\Lambda^2, 36, -6, 72) $ 
in which $S_6$ is the surface area of a $6$-dimensional sphere of 
unit radius and $\Lambda$ is the cut-off for replicated wave-vectors.

Proceeding in the standard way, we find that: 
(i)~For $\wpeps$ negative (i.e.~$d>6$), there is only the 
{\it Gaussian\/} fixed point (GFP),
$(\deltat_*, \inte_*)=(0, 0)$, at which the exponents take on their
classical value: $\nu^{-1}=2$, $\eta=0$. 
(ii)~For $\wpeps$ positive (i.e.~$d<6$), in addition to GFP, a new 
fixed point---the Wilson-Fisher fixed point (WFFP)---emerges, located at
$(\deltat_*,\inte_*^2)=
\left((\Lambda^2/28),(1/126)((2\pi)^6/ S_6)
(V/N)^{-1}\right)\wpeps$, controlling the critical behavior. 
The resulting critical exponents are (to first order in $\wpeps$)
$\nu^{-1}=2-(5\wpeps /{21})$ and $\eta= -{\wpeps/{21}}$.  

A standard scaling argument leads to the value of the critical exponent 
$\beta$ for the gel fraction: 
for $d>6$ we find $\beta=1$; 
for $d<6$ we find $\beta=\nu( d+2-\eta)/2 = 1-\wpeps/7$. 
In fact, under the (not unreasonable) assumption that, in
the ordered state, the fluctuation correlation lengthscale is the same
as the localization lengthscale, we can go further and propose a
general scaling hypothesis for the (singular part of)
$\langle\Omega(\hat k)\rangle^{\FTav}$, viz.,
\begin{equation}
\langle\Omega(\hat k)\rangle^{\FTav}
\sim
\deltat^{\beta}\,w\big({\hat k}^2 \deltat^{2\nu}\big),
\end{equation}
which is supported by the mean-field result~\cite{univ}.  There are, 
however, some fascinating subtleties here, which may instead yield 
multi-fractal characteristics for vulcanized matter similar to those 
explored by Harris and Lubensky in the setting 
of random $xy$ models and resistor networks~\cite{REF:HL1987}.
\par
\noindent
{\it Order-parameter correlator and its physical significance\/}. 
Both above and below six dimensions, the critical exponents $\nu$, 
$\eta$ and $\beta$ of the VT, computed above, have identical values 
(at least to first order in $\wpeps$) to their physical counterparts 
in percolation theory (as computed, e.g., via the Potts field 
theory~\cite{Harris}).  To establish this correspondence, we now 
explore the physical content of the order parameter
correlator~\cite{REF:ZGprior}.  (We remark that the identification 
of the VT $\beta$ with the $\beta$ of percolation is  
evident.)\thinspace\ To proceed, let us consider the construct
\eqbreak
\begin{mathletters}
\begin{eqnarray}
C_{\bft}({\bf r}-{\bf r}^{\prime})
&\equiv&
\bigg[
{V\over{N}}\sum_{j,j^{\prime}=1}^{N}\int_{0}^{1}ds ds^{\prime}
\big\langle
\delta^{(d)}\big({\bf r}-{\bf c}_{j}(s)\big)\, 
\delta^{(d)}\big({\bf r}^{\prime}-{\bf c}_{j^{\prime}}(s^{\prime})\big)\, 
\big\rangle
\big\langle
e^{-i{\bft}\cdot({\bf c}_{j}(s)-{\bf r})}\, 
e^{i{\bft}\cdot({\bf c}_j^{\prime}(s^{\prime})-{\bf r}^{\prime})}
\big\rangle
\bigg]
\label{EQ:PhyDef}
\\
&=& 
{N\over V}
\sum\nolimits_{\bf k}
e^{\rmi({\bf k}+{\bft})\cdot({\bf r}-{\bf r}^{\prime})}
\Big[
\frac{1}{N^2}
\sum_{j,j^{\prime}=1}^{N}
\int_{0}^{1}ds\,ds^{\prime}\,
\big\langle 
e^{-i{\bf k}\cdot
({\bf c}_{j}(s)-
     {\bf c}_{j^{\prime}}(s^{\prime}))}
\big\rangle_\disfac  
\big\langle
e^{-i{\bft}\cdot
({\bf c}_{j}(s)-
     {\bf c}_{j^{\prime}}(s^{\prime}))}
\big\rangle_\disfac  
\Big], 
\end{eqnarray}%
\end{mathletters}%
\eqresume
\noindent
where the term in the second line that is delimited by square brackets 
is the microscopic counterpart of the correlator of a typical 
component of the order parameter, i.e., 
$\big\langle
\Omega({\bf 0},\ldots,{\bf 0},{\bf k},{\bft})^{\ast}\,
\Omega({\bf 0},\ldots,{\bf 0},{\bf k},{\bft})
\big\rangle^{\FTav}$.
The first thermal average in Eq.~(\ref{EQ:PhyDef}) accounts for the
likelihood that monomers $(j,s)$ and $(j^{\prime}, s^{\prime})$ are 
respectively to be found around ${\bf r}$ and ${\bf r}^{\prime}$; the 
second describes the correlation between the respective fluctuations of 
monomer $(j,s)$ about ${\bf r}$ and 
monomer $(j^{\prime},s^{\prime})$ about ${\bf r}^{\prime}$. 

The small-${\bft}$ limit of 
$C_{\bft}({\bf r}-{\bf r}^{\prime})$ addresses the connectedness 
of clusters of mutually crosslinked macromolecules~\cite{REF:smallT}. 
To substantiate this claim, we examine Eq.~(\ref {EQ:PhyDef}), and 
consider the contribution to 
$C_{\bft}({\bf r}-{\bf r}^{\prime})$ from pairs of monomers that 
are in the same cluster and pairs that are in different clusters. 
(We assume that the system has only short-range 
interactions.)\thinspace\ For a generic pair of monomers that are 
in the same cluster, we expect that 
$\langle\exp i{\bft}\cdot
({\bf c}_{j}(s)-{\bf c}_{j^\prime}(s^{\prime}))\rangle \sim 1$, 
and that (for $\vert{\bf r}-{\bf r}^{\prime}\vert\alt R_{\rm g}$)
$\langle\delta^{(d)}\big({\bf r}-{\bf c}_{j}(s)\big)\,
\delta^{(d)}\big({\bf r}^{\prime}-{\bf c}_j(s^{\prime})\big)
\rangle\sim V^{-1}\,R_{\rm g}^{-d}$.  
Then the total {\it intra}-cluster contribution to 
$C_{\bft}({\bf r}-{\bf r}^{\prime})$ is at least of order 
$(N/V)^{2}R_{\rm g}^{-d}$.  
On the other hand, a similar analysis shows that the total 
{\it inter}-cluster contribution is at most of order $(N/V)^{3}V^{-1}$.  
Therefore, the intra-cluster contribution dominates 
$C_{\bft}({\bf r}-{\bf r}^{\prime})$ in the thermodynamic limit. 
(This is also true in the case of generic $\bft$, and thus we 
can always ignore the inter-cluster contribution.)\thinspace\ 
In other words, in the small-${\bft}$ limit, a pair of monomers 
located at ${\bf r}$ and ${\bf r}^{\prime}$ contribute unity
to $C_{\bft}({\bf r}-{\bf r}^{\prime})$ if they are on the same
cluster and contribute zero otherwise.  This limit of 
$C_{\bft}({\bf r}-{\bf r}^{\prime})$ plays the same role as 
the pair-connectedness function defined in (the on-lattice version 
of) percolation theory~\cite{REF:TCL:LH31}.
[For ${\bft}={\bf 0}$, $C_{\bft}({\bf r}-{\bf r}^{\prime})$ is 
simply ($V/N$ times) the real-space density-density correlator, 
and is not of central relevance at the VT.]

In the case of general ${\bft}$, $C_{\bft}({\bf r}-{\bf r}^{\prime})$
addresses the question: If a monomer near ${\bf r}$ is localized on
the scale $\nbft^{-1}$ (or more strongly), how likely is a monomer
near ${\bf r}^{\prime}$ to be localized on the same scale (or more
strongly)? From Eq.~(\ref{EQ:PhyDef}) we see that $\nbft^{-1}$ serves 
as a cutoff to the range of the correlator, so that all pairs of 
monomers that are relatively localized on a length-scale much 
larger than $1/\nbft$ do not contribute to 
$C_{\bft}({\bf r}-{\bf r}^{\prime})$.

Given $C_{\bft}({\bf r}-{\bf r}^{\prime})$, we can build a divergent
susceptibility by integrating 
$C_{\bft}({\bf r}-{\bf r}^{\prime})$ 
over space and  passing to the ${\bft}\to{\bf 0}$ limit: 
\begin{eqnarray}
&&\lim_{{\bft}\to{\bf 0}}
\int
\frac {\rmd^{d}r\,\rmd^{d}r^{\prime}}{V}\,
C_{\bft}({\bf r}-{\bf r}^{\prime})\sim
\\
&&
N\lim_{n \to 0}
\left\langle
\Omega({\bf 0},..,{\bf 0},{\bft},-{\bft})^{\ast}\,
\Omega({\bf 0},..,{\bf 0},{\bft},-{\bft})
\right\rangle^{\FTav}
\sim 
(-\deltat)^{-\gamma}.
\nonumber
\label{EQ:SusceptDef}
\end{eqnarray}
This quantity is measure of the spatial extent over which pairs of 
monomers are relatively localized, no matter how weakly, and thus 
diverges at the VT.   
\par
\noindent
{\it Concluding remarks\/}. 
As we have seen, the VT exponents obtained by the $\wpeps$ expansion 
turn out to be numerically equal to those characterizing physically
analogous quantities in percolation theory, at least to first order 
in $\wpeps$.  This equality between exponents seems reasonable, in 
view of the intimate relationship between percolation theory and the 
{\it connectivity\/} of the system of crosslinked macromolecules.  
Such a  connection has long been recognized, and supports the use of 
percolative approaches as models of certain aspects of the
VT~\cite{REF:FloryBook,REF:PGDGbook,REF:LubIsaac}.

A convenient formulation of the percolative approach is via the Potts
model in its one-state limit~\cite{REF:FortKast}.  It is therefore
worth considering similarities and differences between the minimal
model of VT, Eq.~(\ref{EQ:LG_longwave}), and the minimal field theory
for the Potts model.  Both have a cubic interaction and both involve
an $n \to 0$ limit. Despite the similarities, however, there exist 
many differences:
(i)~The Potts field theory has a multiplet of $n$ real fields on
$d$-dimensional space; the VT field theory has a real singlet field
living on $(n+1)d$-dimensional space.
(ii)~ The Potts field theory emerges from a setting involving a 
{\it single\/} ensemble, whereas the VT field theory describes a 
physical problem in which {\it two\/} distinct ensembles (thermal 
and disorder) play essential roles.  As such, the vulcanization 
field theory is a more natural and direct approach, in 
the sense that it is capable 
of providing a unified theory of both the transition and the 
structure of the solid state.
(iii)~The symmetry structures are different; most strikingly,
the nature of the spontaneous symmetry breaking is different.
The percolation transition (in its Potts representation) involves the
spontaneous breaking of the ($n\to 0$ limit) of a {\it discrete\/} 
$(n+1)$-fold permutation symmetry.  
By contrast, the VT involves the spontaneous breaking of the ($n\to 0$
limit of the) {\it continuous\/} symmetry of relative translations and
rotations of the $n+1$ replicas; the permutation symmetry remains
intact in the amorphous solid state, as does the symmetry of common
translations and rotations of replicated space.
The fact that the critical exponents near the upper critical dimension
are the same for both theories (at least to order $\wpeps$) despite 
these apparent distinctions can be reconciled by delineating between 
three logically distinct physical properties pertaining to amorphous
solidification transition:
(i)~macroscopic network formation;
(ii)~changes in the nature of thermal 
motion---i.e.~random localization; and 
(iii)~the acquisition of rigidity.
Within mean-field theory (and hence above six spatial dimensions),
these three properties go hand-in-hand, emerging simultaneously at 
the VT.  At and below six dimensions they appear to continue to 
go hand-in-hand (although we have not yet investigated the issue 
of rigidity beyond mean-field theory) until one reaches two 
dimensions, where (as we shall discuss shortly) we believe this 
broad picture changes.  Thus, in higher dimensions it appears 
that when both ensembles are incorporated the disorder fluctuations 
play a more important role than the thermal fluctuations.

This brings up the interesting issue of the nature of 
the VT at the neighborhood of two dimensions---the lower 
critical dimension of the VT~\cite{REF:stability}. 
(The ideas reported in this paragraph result from an ongoing
collaboration with H.~E.~Castillo~\cite{REF:withHEC}.)\thinspace\
Indeed, there is a conventional percolation transition in two
dimensions, whereas the thermal fluctuations are expected to
destroy the random localization.  It is tempting to
speculate~\cite{REF:withHEC} that in two dimensions an anomalous type
of VT happens simultaneously with percolation transition: As the
constraint-density is tuned from below to above its critical value, 
the order parameter would remain zero, whereas its correlations would 
change from decaying exponentially to decaying algebraically: one would
have a quasi-amorphous solid state---the random analog of a
two-dimensional solid~\cite{REF:DRN:DandG}.  
\par
\noindent
{\it Acknowledgments\/}.
It is a pleasure to thank 
K.~Dahmen, 
B.~Fourcade, 
E.~Fradkin, 
S.~Glotzer, 
A.~Halperin, 
T.~Lubensky, 
J.~M.~Rom\'an, 
M.~Stone, 
C.~Yu, and especially 
H.~Castillo 
for helpful discussions.  
This work was supported by 
NSF DMR99-75187 (WP, PMG) and 
    DMR91-20000 (WP).
  
\end{multicols}

\begin{references}
\bibitem{REF:smallT}
By small $\vert\bft\vert$ we mean 
$ R_{\rm g} \ll \vert\bft\vert^{-1} \ll V^{-1/3} $, 
with $R_{\rm g}$ being the radius of gyration 
for the largest cluster.
\end{references}
\end{document}